\documentstyle[preprint,aps]{revtex}
\tightenlines

\def\cm{~cm$^{-1}$}
\def\tc{T${_{\mathrm{c}}}$}
\def\et{{\it et al.}}
\def\ybco{YBa$_2$Cu$_3$O$_7$}
\def\square{\kern1pt\vbox{\hrule height 0.6pt\hbox{\vrule width 
0.6pt\hskip 2pt \vbox{\vskip 4pt}\hskip 2pt\vrule width 0.6pt}\hrule
height 0.6pt}\kern1pt}
\def\h{_{\mathrm{H}}}
\def\f{_{\mathrm{F}}}
\def\xx{_{\mathrm{xx}}}
\def\xy{_{\mathrm{xy}}}

\begin{document}

\title{Infrared Hall effect in high \tc\ superconductors: Evidence for
non-Fermi liquid Hall scattering}

\author{J. \v Cerne,$^1$ M. Grayson,$^1$ D.C. Schmadel,$^1$ G.S.
Jenkins,$^1$ H.D. Drew,$^1$ R. Hughes,$^2$ J.S. Preston,$^2$ and P.-J.
Kung$^3$}

\address{$^1$Center for Superconductivity Research and Department of
Physics, University of Maryland, College Park, MD 20741, USA.}

\address{$^2$ Department of Physics and Astronomy, McMaster University,
Hamilton, ON L8S 4M1, Canada.}

\address{$^3$Advanced Fuel Research, Inc., East Hartford, CT 06108, USA.}

\maketitle

\begin{abstract}
Infrared (20-120\cm\ and 900-1100\cm) Faraday rotation and circular
dichroism are measured in high \tc\ superconductors using sensitive
polarization modulation techniques. Optimally doped \ybco\ thin films are
studied at temperatures down to 15~K and magnetic fields up to 8~T. At
1000\cm\ the Hall conductivity $\sigma\xy$ varies strongly with
temperature in contrast to $\sigma\xx$ which is nearly independent of
temperature. The Hall scattering rate $\gamma\h$ has a $T^{2}$ temperature
dependence but, unlike a Fermi liquid, depends only weakly on frequency.
The experiment puts severe constraints on theories of transport in the
normal state of high \tc\ superconductors.
\end{abstract}


\date{8/2/99}

\narrowtext


The ``normal'' state of high \tc\ superconductors (HTSC) shows behaviors
in many physical properties that are anomalous in comparison with
conventional metals \cite{andersonbk}. These anomalous properties occur in
both the optimally doped materials that we discuss in this Letter and even
more dramatically in the underdoped pseudogap materials. The DC
resistivity $\rho$ varies linearly with temperature $T$ from \tc\ to the
melting point. The Hall coefficient $R\h$ while positive and decreasing
with hole doping, varies with temperature as $1/T$ \cite{kamaras}. The
cotangent of the DC Hall angle $\theta\h,$ which in simple metals has the
same temperature dependence as $\rho$, varies in HTSC as $T^{2}$
\cite{chien1}. This anomalous behavior of the Hall effect appears to occur
in the superconducting state as well, where thermal magneto-transport
shows a similar $T^{2}$ behavior of the thermal Hall angle and thermal
Hall scattering rate below \tc\ \cite{zeini}.  In contrast, the zero field
quasiparticle scattering rate in the superconducting state has a $T^{4}$
dependence \cite{bonn}. The normal state anomalies have been observed in
DC measurements over a wide range of temperatures and in a wide variety of
HTSC materials. Normal state far infrared (far-IR) measurements near \tc\
have determined that the scattering rate associated with $\theta\h$ is 3-4
times smaller than the scattering rate associated with the longitudinal
conductivity $\sigma\xx$ \cite{kaplan}. On the other hand, recent
angularly resolved photoemission spectroscopy (ARPES) work reports that
the quasiparticle scattering rate $\gamma$ is minimum along the $(\pi ,\pi
)$ direction of the Brillouin zone and varies as $\gamma \approx \max
(\omega,\pi T)$ \cite{arpes}.  Also, since the Fermi velocity is maximum
in this direction these quasiparticles should dominate both $\sigma\xx$
and $\sigma_{{\rm xy}}$ \cite{arpes}. Therefore experiments not only
suggest that the longitudinal conductivity $\sigma\xx$ and the transverse
conductivity $\sigma\xy$ differ fundamentally in HTCS but that they differ
from the predictions of Fermi liquid theory.

There have been many different proposed explanations for the anomalous
Hall effect in HTSC \cite{carrington,ioffe,zhel,kotliar,anderson,coleman}.
These explanations can be classed into two basic theoretical approaches.
The first approach argues that the system is a Fermi liquid (FL), with
excitations consisting of hole-like quasiparticles, but that the strong
anisotropy of the Fermi surface and/or the quasiparticle scattering on the
Fermi surface causes a non-Drude behavior of the transport \cite
{carrington,ioffe,zhel}. The other theoretical approaches argue that the
system has non-FL properties, with excitations composed of more exotic
entities. In Anderson's theory \cite{anderson} based on Luttinger liquid
ideas spinons and holons have different relaxation mechanisms. Coleman et
al. \cite{coleman} have considered the transport processes of the
quasiparticles associated with charge conjugation symmetry of the system.  
Many of these models can account qualitatively for the DC measurements.
This circumstance provides a motivation for extending measurements into
the infrared (IR), where experiments may allow a critical test of the
proposed models.

In this Letter we report measurements of the magneto-optical response of
optimally doped \ybco\ thin films in a broad range of far-IR (20-140 \cm)
and mid-infrared (mid-IR, 900-1100 \cm) frequencies using novel
polarization modulation techniques \cite{au}. From these measurements and
the zero field optical conductivity we extract the full
magneto-conductivity as well as the complex Hall angle $\theta\h$.

The samples are optimally doped \ybco\ thin films grown on Si or
LaSrGaO$_{4}$ substrates.  The primary sample used for the mid-IR
measurements consists of a 150~nm thick film grown on LaSrGaO$_4$, with a
\tc\ and $\Delta$\tc\ of 88.2~K and 0.6~K, respectively. Since
interference (etalon) effects can have a strong effect on IR Hall
measurements, the substrates were either coated with a NiCr broadband
antireflection coating \cite{ar} or wedged 1 degree to remove multiply
reflected beams.

The measured quantity in the magneto-optical experiments is the complex
Faraday angle $\theta _{F}$ defined as $\tan \theta\f=t\xy/t\xx$, where
$t\xx$ and $t\xy$ are the complex transmission amplitudes. By determining
$\sigma\xx$ through zero magnetic field transmittance and reflectance
measurements, Maxwell's equations can be used to transform $\theta\f$ into
the more interesting transport quantities, $\sigma\xy$, the off-diagonal
component of the complex magneto-conductivity tensor, $\theta\h$, the
complex Hall angle defined as \cite{au} $\tan
\theta\h=\sigma\xy/\sigma\xx$ and the complex Hall coefficient
$R\h=\sigma\xy/\sigma\xx^2$. For the experiments reported in this Letter,
$\theta\h$ is small, so that $\tan \theta\f\approx \theta\f$ and
$\tan\theta\h\approx \theta\h$. $\theta\h$ has the analytic properties of
a response function, obeying Kramers-Kronig relations (KKR) and a sum rule
\cite{drew}. In general $\tan \theta\h$, the ratio of two response
functions, is a complicated function which does not have a simple closed
form. The simplest generalization of $\theta\h^{-1}$ to finite frequency
$\omega $ is \cite{achall,au}:

\begin{equation}
\theta\h^{-1}=\frac{\gamma\h}{\omega\h}-i\frac{\omega }{\omega\h},  
\label{eq;invtheta}
\end{equation}

\noindent where $\omega\h$ is the Hall frequency and $\gamma\h$ is the
Hall scattering frequency. Equation~\ref{eq;invtheta} is valid for a Drude
metal in which case, $\omega\h=\omega _{c}$ and $\gamma\h=\gamma$, where
$\omega _{c}$ and $\gamma $ are the conventional cyclotron frequency and
isotropic Drude scattering rate, respectively. Equation~\ref{eq;invtheta}
is also valid for a Fermi liquid for the case of a $k$ independent
scattering time, and it is the form obtained in several proposed models of
the normal state transport in HTSC \cite{zhel,ioffe,anderson}.  
Furthermore, Eq.~\ref{eq;invtheta} represents the general high frequency
limiting behavior of $\theta\h$ \cite{au}.


In the far-IR we measure the Faraday rotation with a rotating linear
polarizer. We fit the data to the empirically observed Lorentzian behavior
of the Hall angle \cite{kaplan} (see Eq.~\ref{eq;invtheta}). The Hall
frequency, $\omega\h$, and Hall scattering rate, $\gamma\h=1/\tau\h$ are
determined from the fit.

In the mid-IR, the Hall angle is small (on the order of $10^{-3}$ Rad at
8~T) since $\omega\h<<(\gamma\h,\omega)$ for this experiment. Therefore, a
sensitive technique is required for the mid-IR $\theta\h$ measurements
\cite{au}.  A CO$_2$ laser produces linearly polarized mid-IR radiation
which passes in the Faraday geometry through the sample, located at the
center of an 8~T magneto-optical cryostat. In order to sensitively measure
both the real and imaginary parts of $\theta\f$ (i.e., the rotation and
ellipticity of the transmitted polarization), the radiation that is
transmitted by the sample is analyzed using a ZnSe photoelastic modulator
(PEM).  A liquid nitrogen-cooled mercury-cadmium-telluride element detects
the radiation, and three lock-in amplifiers demodulate the resulting
time-dependent signal. By analyzing both the even and odd harmonic signals
of the PEM modulation frequency, one can simultaneously measure both the
real and imaginary parts of $\theta\f$ with a sensitivity of better than 1
part in $10^{4}$ and $4\times 10^{3}$, respectively.


Figure~\ref{fig;hallcond} shows the complex Hall conductivity $\sigma\xy$
as a function of temperature at 8~T. The Hall conductivity is the most
directly accessible transport response function determined from the
experiment. It is also interesting because it is expected to be least
affected by the conductivity of the Cu-O chains.  Re[$\sigma\xy$] in
Fig.~\ref{fig;hallcond}(a) shows strong temperature dependence whereas the
Im[$\sigma\xy$] in Fig.~\ref{fig;hallcond}(b) shows little temperature
dependence. The factor of five increase in Re[$\sigma\xy$] at low
temperature is striking since both the real and imaginary parts of
$\sigma\xx$ at 1000\cm\ vary by less than 20~\% over the same temperature
range \cite{schlesinger,bomem}. Note that the Re[$\sigma\xy$] only shows
frequency dependence at lower temperatures while Im[$\sigma\xy$] shows a
uniform decrease with frequency across all temperatures. $\sigma\xy$ is
not consistent with a Drude model with a temperature dependent carrier
scattering rate.  Nor is it in the high frequency limit $\left( \omega \gg
\gamma \right) $ where general considerations imply that $\sigma\xy\propto
\left( -1/\omega ^{2}+2i\gamma /\omega ^{3}\right)$, where $\gamma $ is a
momentum space average of the scattering rate \cite{achall,au}. The weak
frequency dependence of Re[$\sigma\xy$] and the fact that Im[$\sigma\xy$]
is not small compared with Re[$\sigma\xy$] indicates that by 1000\cm\ the
system is not yet in the asymptote $\left( \omega >\gamma \right)$ regime.

We have also examined the complex Hall angle and the Hall coefficient.
Since \ybco\ contains Cu-O chains in addition to the Cu-O planes, the
longitudinal conductivity $\sigma\xx$ is anisotropic in single domain
samples \cite{chain1,chain2}. Therefore, the $\sigma\xx$ measured for the
twinned thin films used in this experiment is an average of the
conductivities of the chains and planes. It is most interesting to examine
the transport quantities of the Cu-O planes since the chains do not
contribute significantly to the superconductivity.  We can use the results
for $\sigma\xx$ on single domain samples in Ref.~\cite{chain2} to
characterize the chain conductivity in \ybco.  The chain conductivity is
sample dependent so that we can only estimate their effects on our twinned
films. We expect the chain contribution to be smaller than observed in
single domain samples since our polycrystalline films are more likely to
have disorder which can very easily upset the one dimensional chain
conductivity.  This is confirmed by our measurements of the mid-IR
$\sigma\xx$, whose temperature and frequency behavior is consistent with a
conductivity that is dominated by the planes.  We note that the chain
contribution in the far-IR has been reported to be negligible
\cite{chain1,chain2}.  From these considerations we conclude that the
magnitude of the chain corrections is less than 30~\% and 10~\% for the
real and imaginary parts of the mid-IR $\sigma\xx$ respectively. Because
of the uncertainties, however, we have not removed the contribution from
the Cu-O chains in this Letter but rather we will discuss their effects as
we present the data.

Because of the form of $\theta\h$ given by many of the theoretical models
(see Eq.~\ref{eq;invtheta}) it is most interesting to examine the complex
inverse Hall angle $\theta\h^{-1}$. Figure~\ref{fig;invhall} shows the
temperature dependence of the mid-IR $\theta\h^{-1}$ at 8~T. The
Re[$\theta\h^{-1}$] shows strong temperature dependence in
Fig.~\ref{fig;invhall}(a) consistent with a temperature dependent
$\gamma\h$. The Im[$\theta\h^{-1}$] is nearly temperature independent in
the normal state in Fig.~\ref{fig;invhall}(b), which is consistent with a
nearly temperature independent $\omega\h$ (see Eq.~\ref{eq;invtheta}). The
solid line in Fig.~\ref{fig;invhall}(a) shows the measured DC
$\theta\h^{-1}$ which is seen to agree well with the mid-IR
Re[$\theta\h^{-1}$]. The Re[$\theta\h^{-1}$] in Fig.~\ref{fig;invhall}(a)
shows no frequency dependence, which is consistent with a frequency
independent $\gamma\h$, while the frequency dependence of
Im[$\theta\h^{-1}$] is consistent with Eq.~\ref {eq;invtheta} and a
frequency independent $\omega\h$.

If we assume the Lorentzian form for $\theta\h$ given by
Eq.~\ref{eq;invtheta} we can extract the normal state Hall frequency
$\omega\h$ and Hall scattering rate $\gamma\h$ which are shown in
Fig.~\ref{fig;hallscat} as a function of temperature at 8~T. $\omega\h$
shows little temperature dependence in Fig.~\ref{fig;hallscat}(a), while
strong temperature dependence is observed for the mid-IR $\gamma\h$ in
Fig.~\ref{fig;hallscat}(b). Both $\omega\h$ and $\gamma\h$ appear to be
frequency independent. Also, both the mid-IR $\omega\h$ and $\gamma\h$ are
in good with agreement our far-IR measurements and with the 200\cm\ result
obtained by Ref.~\cite{kaplan}.  The agreement of the far-IR and mid-IR
$\omega\h$ improves when the estimated Cu-O chain contribution to
$\sigma\xx$ is removed, which causes an increase of up to 30~\% in the
mid-IR $\omega\h$. This correction only weakly affects the mid-IR
$\gamma\h$, which is reduced by less than 10~\%.

The previous results are used to determine the mid-IR complex Hall
coefficient $R\h$ as a function of temperature.
For a Drude metal $R\h$ is real and independent of frequency. While the
measured Re[$R\h$] shows weak temperature dependence the Im[$R\h$] is
large and shows strong temperature dependence. At low temperatures, $R\h$
is mostly imaginary showing again that the experiment is not near the high
frequency limit at 1000\cm. The DC values of $R\h$ are more than a factor
of five larger than the mid-IR value at 100~K, but approach the mid-IR
results at higher temperatures.  No frequency dependence is observed for
the mid-IR Re[$R\h$], while the mid-IR Im[$R\h$] is large and weakly
frequency dependent at lower temperatures.

The agreement of the mid-IR and far-IR results in Fig.~\ref{fig;hallscat}
shows that the frequency dependence of $\omega\h$ and $\gamma\h$ is very
weak. The conductivity relaxation rate $\gamma $ for a Fermi liquid has
the following form:

\begin{equation}
\gamma=\frac{1}{W}\biggl[ \biggl( \frac{\omega}{p\pi}\biggr)^2+T^2\biggr],  
\label{eq;flgamma}
\end{equation}

\noindent where $p=2$ is the calculated result \cite{gurzhi} for the
optical response and $p=1$ is observed for heavy fermion systems
\cite{sievers}.  $W$ represents a characteristic energy scale for the
electron system. We find $W\approx 120$~K from the far-IR data
\cite{kaplan,grayson}.  The mid-IR data fits a $T^2$ dependence (thin
solid line in Fig.~\ref{fig;hallscat}(b)) with a small offset. This
results in an estimate for $W$ that is approximately 160~K.  The predicted
$\gamma\h$ based on Fermi liquid theory and the observed $T^{2}$
dependence of $\gamma\h$ in the IR is shown as a dashed line in
Fig.~\ref{fig;hallscat}(b) which clearly disagrees with the mid-IR
results, demonstrating the weak frequency dependence and a non-Fermi
liquid behavior of the Hall scattering.

The weak frequency dependence of $\gamma\h$ is internally consistent with
the agreement of the far-IR and mid-IR values for $\omega\h$. As discussed
previously \cite{achall}, a frequency dependent $\gamma\h$ should produce,
by the properties of response functions, a frequency dependent effective
$\omega\h$ saturating to its true sum rule value as $\omega \rightarrow
\infty $. It was pointed out in the earlier work \cite {achall} that
$\omega\h(\omega )$ for both the superconducting ($\omega\h^{{\rm s}}$)
and normal state ($\omega\h^{{\rm n}}$)  {\it appeared} to saturate by
200\cm, but they saturated to different values with $\omega\h^{{\rm
s}}(200)\simeq 2\omega\h^{{\rm n}}(200)$.  The mid-IR data show that
$\omega\h^{n}(200)$ is close to the correct sum. Therefore it is expected
that Re[$\theta\h$] in the superconducting state must be negative between
200\cm\ and 1000\cm. This is just the behavior that the superconducting
energy gap in $\sigma\xx$ is expected to cause and is confirmed by the
mid-IR measurement, which shows that Re[$\theta\h$] becomes negative below
\tc.

These IR Hall results can be compared with theoretical models of the Hall
conductivity. The Ong-Anderson model \cite{anderson} assumes that
$\sigma\xx$ and $\theta\h$ are controlled by different scattering rates.
The IR experiments can be explained within this model by choosing the
frequency and temperature dependence of these rates appropriately. In the
two scattering time ($\tau$) models of Refs.~\cite{zhel} and
\cite{coleman}, the two $\tau$'s do not separately control $\sigma\xx$ and
$\theta\h$ so that the experimentally observed behavior does not occur so
naturally. The skew scattering model \cite{kotliar} fails because it
predicts the wrong temperature and frequency dependence of the far-IR
magneto-conductivity \cite{zhel}. The Ioffe-Millis model \cite{ioffe}
involves only one $\tau$.  Alhough this model is consistent with the
mid-IR data, it appears to contradict the two $\tau$ behavior observed in
the far-IR \cite{zhel2}.  More detailed analysis of data on several
different HTSC materials may allow definitive tests of these models.

The Hall scattering rate obtained in this work has a $T^{2}$ temperature
dependence but is independent of frequency. This behavior differs from the
IR conductivity and the ARPES results in high \tc\ superconductors which
show that the quasiparticle scattering rate has a linear dependence on
both temperature and frequency. In general, the temperature dependence of
carrier relaxation in solids, which usually arises from inelastic
scattering from phonons, magnons, or the electron-electron interaction,
has a frequency dependence similar to the temperature dependence.  
Therefore the frequency and temperature dependence of $\gamma\h$ that are
reported in this Letter are highly unusual and indicate non-Fermi liquid
behavior of the normal state of \ybco. However, Ioffe and Millis
\cite{ioffe} have recently proposed such a relaxation rate behavior based
on quasi-elastic scattering from superconducting fluctuations in the
normal state of high \tc\ materials. Fluctuation effects have also been
observed in normal state of underdoped Bi$_2$Sr$_2$CaCu$_2$O$_8$
in measurements of the longitudinal conductivity by THz spectroscopy
\cite{orenstein}.


We acknowledge useful discussions with A.J. Millis, N.P. Ong, V. Yakovenko
and P.D. Johnson. This work was supported in part by NSF grant DMR-9705129
and by the NSA.

\begin{figure}[tbp]
\caption{The real (a) and imaginary (b) parts of the Hall conductivity
$\sigma\xy$ at 8~T as a function of temperature at 949\cm\ ($\bullet$) and
1079\cm\ ($\triangle$). }
\label{fig;hallcond}
\end{figure}

\begin{figure}[tbp]
\caption{The real (a) and imaginary (b) parts of the inverse Hall angle
$\theta\h^{-1}$ at 8~T as a function of temperature at 949\cm\ ($\bullet$)
and 1079\cm\ ($\triangle$). The thin solid line in (a) shows the values
for $\theta\h^{-1}$ obtained using DC Hall measurements at 8~T.}
\label{fig;invhall}
\end{figure}

\begin{figure}[tbp]
\caption{(a) The Hall frequency, $\omega\h$, and (b) the Hall scattering
rate, $\gamma\h$, at 8~T as a function of temperature for far-IR (\square=
20-150\cm, and $+$ is from Ref.~\protect\cite{drew}) and mid-IR ($\bullet=
949$\cm, and $\triangle = 1079$\cm) frequencies. The solid line in (b)  
shows a $T^2$ fit to the mid-IR data, with the dashed line showing the
expected mid-IR Hall scattering rate based on Fermi liquid theory
(Eq.~\protect\ref{eq;flgamma}). Equation~\protect\ref{eq;invtheta} ceases
to be relevant in the hatched region below \tc, but the data are plotted
for completeness. }
\label{fig;hallscat}
\end{figure}


\begin{references}
\bibitem{andersonbk}  P.W. Anderson, {\it The Theory of Superconductivity in
the High-\tc\ Cuprates} (Princeton University Press, Princeton,
1997).

\bibitem{kamaras}  K. Kamaras \et, Phys. Rev. Lett. {\bf 64}, 84
(1990), and  L. Forro \et, Phys. Rev. Lett. {\bf 65}, 1941
(1990).

\bibitem{chien1} T.R. Chien \et, Phys. Rev. B. {\bf 43}, 6249 (1991), T.R.
Chien \et, Phys. Rev. Lett. {\bf 67}, 2088 (1991), and J.M. Harris \et,
Phys. Rev. B. {\bf 46}, 14293 (1992).

\bibitem{zeini}  Zeini \et, Phys. Rev. Lett. {\bf 82}, 2175 (1999).

\bibitem{bonn}  A. Hosseini \et, Phys. Rev. B {\bf 60}, 1349 (1999).

\bibitem{kaplan} S.G. Kaplan \et, Phys. Rev. Lett. {\bf 76}, 696 (1996).

\bibitem{arpes} T. Valla \et, submitted to Science, 1999, and A. Kaminski
\et, cond-mat/9904390, 1999.

\bibitem{carrington} A. Carrington \et, Phys. Rev. Lett. {\bf 69}, 2855
(1992).

\bibitem{ioffe} L.B. Ioffe and A.J. Millis, Phys. Rev. B {\bf 58}, 11631
(1998).

\bibitem{zhel} A.T. Zheleznyak \et, Phys. Rev. B {\bf 57}, 3089 (1998).

\bibitem{kotliar} G. Kotliar \et, Phys. Rev. B {\bf 53}, 3573 (1996).

\bibitem{anderson} P.W. Anderson, Phys. Rev. Lett. {\bf 67}, 2092 (1991).

\bibitem{coleman}  P. Coleman \et, Phys. Rev. Lett. {\bf 76}, 1324 (1996).

\bibitem{au}  J. \v Cerne \et, cond-mat/9907210.

\bibitem{ar} S.W. McKnight \et, Infrared Phys. {\bf 27}, 327 (1987).

\bibitem{drew} H.D. Drew and P. Coleman, Phys. Rev. Lett. {\bf 78}, 1572
(1997).

\bibitem{achall} H.D. Drew \et, J. Phys.: Condens. Matter {\bf 8}, 10037
(1996).

\bibitem{schlesinger} Z. Schlesinger \et, Phys. Rev. B {\bf 41}, 11237
(1990).

\bibitem{bomem} The temperature dependence of the complex $\sigma\xx$ was
measured for the sample used in this experiment by performing zero
magnetic field transmittance and reflectance measurements.

\bibitem{chain1} D.N. Basov, Phys. Rev. Lett. {\bf 74}, 598 (1995), and
S.L. Cooper \et, Phys. Rev. B {\bf 47}, 8233 (1993).

\bibitem{chain2} J. Sch\"{u}tzmann \et, Phys. Rev. B {\bf 46}, 512 (1992).

\bibitem{gurzhi} R.N. Gurzhi \et, Zh. Eksp. Teor. Fiz. {\bf 35}, 965
(1958)  [Sov. Phys.-JETP {\bf 8}, 673 (1959).

\bibitem{sievers} P.E. Sulewski \et, Phys. Rev. B {\bf 38}, 5338 (1988).

\bibitem{grayson}  M. Grayson \et, in preparation.

\bibitem{zhel2}A.T. Zheleznyak \et, Phys. Rev. B {\bf 59}, 207 (1999).

\bibitem{orenstein}J. Colson \et, Nature {\bf 398}, 22 (1999).
\end{references}
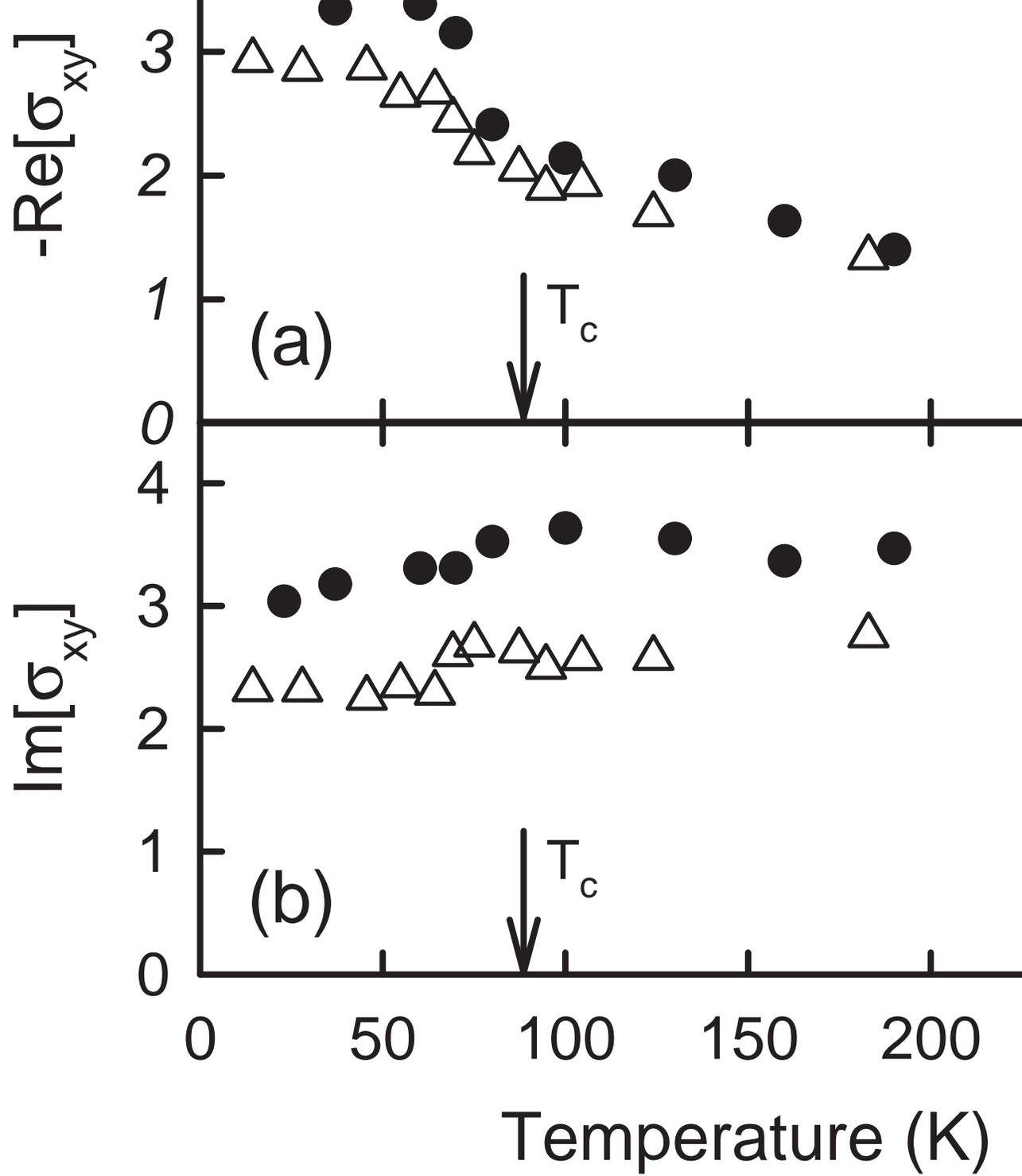
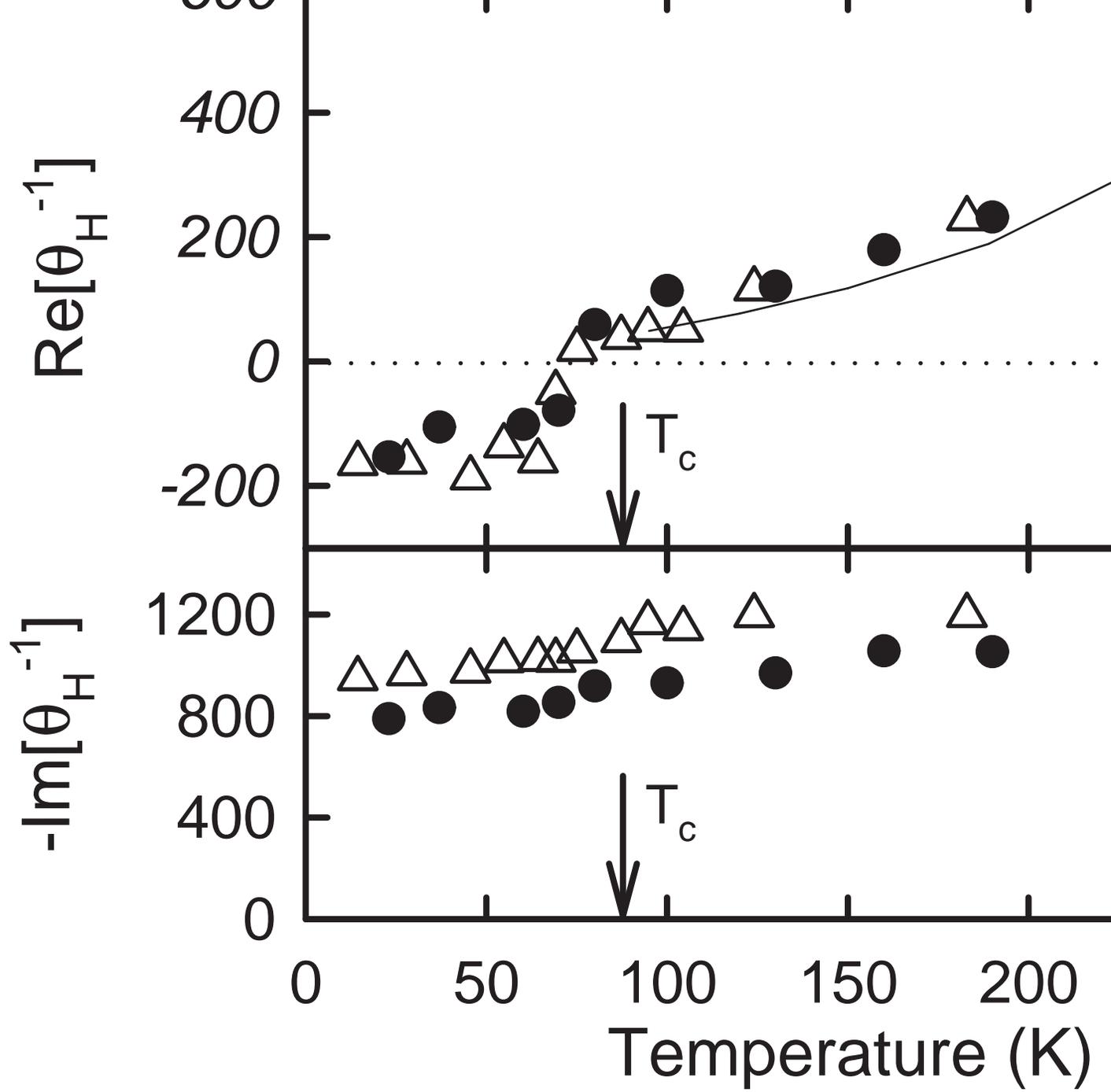
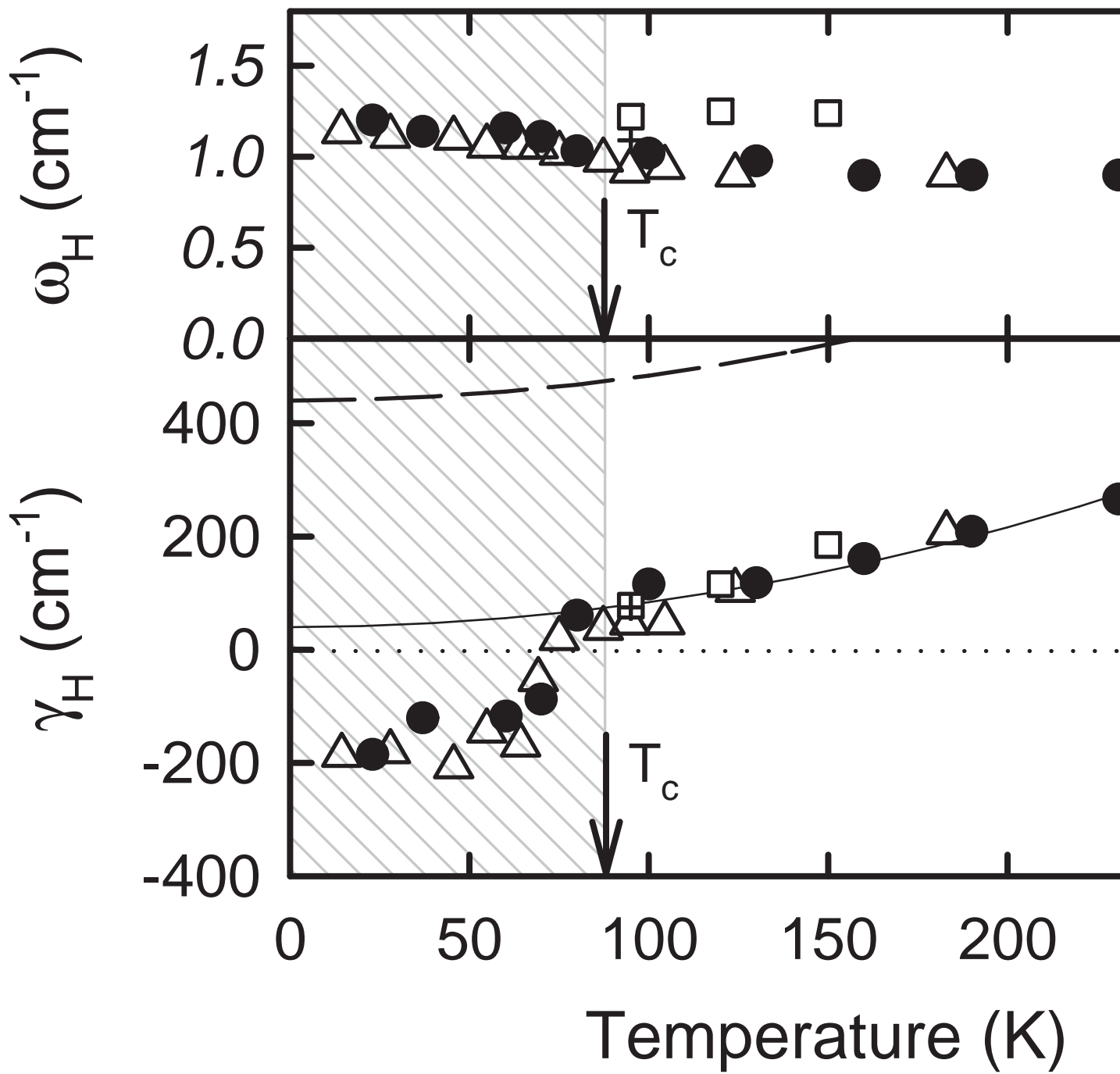
\end{document}